# Pandemics are catalysts of scientific novelty: Evidence from COVID-19

Meijun Liu[1], Yi Bu[2], Chongyan Chen[3], Jian Xu[4], Daifeng Li[4], Yan Leng[5], Richard Barry Freeman[6,7], Eric Meyer[3], Wonjin Yoon[8], Mujeen Sung[8], Minbyul Jeong[8], Jinhyuk Lee[8], Jaewoo Kang[8,9], Chao Min[10], Min Song[11], Yujia Zhai[12,13], Ying Ding [4,14*]

[1]Institute for Global Public Policy, Fudan University, Shanghai, China.

[2] Department of Information Management, Peking University, Beijing, China.

[3]School of Information, University of Texas at Austin, Austin, TX, USA.

[4]School of Information Management, Sun Yat-sen University, Guangzhou, China.

[5]McCombs School of Business, University of Texas at Austin, Austin, TX, USA.

[6]Department of Economics, Harvard University, Cambridge, MA, USA.

[7]National Bureau of Economic Research (NBER), Cambridge, MA, USA.

[8]Department of Computer Science and Engineering, Korea University, Seoul, Korea.

[9]Interdisciplinary Graduate Program in Bioinformatics, Korea University, Seoul, Korea.

[10] School of Information Management, Nanjing University, Nanjing, China.

[11]Department of Library and Information Science, Yonsei University, Seoul, Korea.

[12]Department of Information Resource Management, School of Management, Tianjin Normal University, Tianjin, China.

[13]School of Information Management, Wuhan University, Wuhan, China.

[14]Dell Medical School, University of Texas at Austin, Austin, TX, USA.

*Corresponding author: E-mail: ying.ding@austin.utexas.edu

## Abstract

Scientific novelty drives the efforts to invent new vaccines and solutions during the pandemic. First-time collaboration and international collaboration are two pivotal channels to expand teams' search activities for a broader scope of resources required to address the global challenge, which might facilitate the generation of novel ideas. Our analysis of 98,981 coronavirus papers suggests that scientific novelty measured by the BioBERT model that is pre-trained on 29 million PubMed articles, and first-time collaboration increased after the outbreak of COVID-19, and international collaboration witnessed a sudden decrease. During COVID-19, papers with more first-time collaboration were found to be more novel and international collaboration did not hamper novelty as it had done in the normal periods. The findings suggest the necessity of reaching out for distant resources and the importance of maintaining a collaborative scientific community beyond nationalism during a pandemic.

# Introduction

Scientific novelty advances knowledge frontier and drives technological innovation. One of the key issues in science of science is how scientific novelty origins and develops (Fortunato et al., 2018; Uzzi, Mukherjee, Stringer, & Jones, 2013). Driven by the outbreak of COVID-19, a particular issue of interest is the evolution of scientific novelty during unexpected crises beyond a more conventional scientific environment. The importance of scientific novelty became more salient during COVID-19 since the key to attacking COVID-19 and recovering from the aftermath of the pandemic lies in finding innovative and effective solutions (Azoulay & Jones, 2020; El Akoum & El Achi, 2021a).

Despite the importance of scientific novelty, it remains unclear whether and how scientific novelty evolved during COVID-19. Extensive studies have documented the detrimental effects of COVID-19 on scientists in various aspects, ranging from a decline in working hours (Myers et al., 2020), increasing difficulties in collaboration (Aviv-Reuven & Rosenfeld, 2021; Cai, Fry, & Wagner, 2021) to reduction in initiating new projects (Gao, Yin, Myers, Lakhani, & Wang, 2021b). These negative impacts might dampen scientists' capacities to innovate. However, some believe that crises could be drivers of innovation due to the urgency for addressing the unprecedented challenges and the need for fast solutions to new problems (Birkland, 2004; Gopalakrishnan & Kovoor-Misra, 2021; Harris, Bhatti, Buckley, & Sharma, 2020; Knudsen, 2019). The stunning advancement of vaccines at an amazing speed[1] exemplified that COVID-19 might create a fertile breeding ground for scientific novelty[2] and have the potential to be an innovator trigger (Kim, Dema, & Reyes-Sandoval, 2020; Ramalingam & Prabhu, 2020). Despite anecdotal evidence that claims accelerated innovation processes during COVID-19, empirical evidence is still absent in the literature. Therefore, one overarching question arises: **RQ1. How did scientific novelty evolve during COVID-19?**

Facing increasing resources constraints during COVID-19, scientists were involved in teams to complement resources they could access for generating novel solutions (Cai et al., 2021; Fry, Cai, Zhang, & Wagner, 2020a; Wagner, Cai, Zhang, & Fry, 2021). The lack of time and resources available for effective and timely responses made it difficult to push coronavirus research forward by either individual or intra-country efforts (Fry, Cai, Zhang, & Wagner, 2020b). The sharing of knowledge, data and other resources became more essential than ever, which required collaboration among scientists, organizations and countries (Cai et al., 2021; Wagner et al., 2021). Additionally, the information processing model also argues that scientists could access a broader scope of information and diverse ideas by collaborating with others and thus produce more novel ideas (Harrison & Klein, 2007; Lee, Walsh, & Wang, 2015; Nederveen Pieterse, Van Knippenberg, & Van Dierendonck, 2013).

The evolution of scientific novelty during COVID-19 might be accompanied by changes in its influential factors, especially collaboration-related factors, due to the dominance of teams in the production of knowledge (Wuchty, Jones, & Uzzi, 2007).[3] Scientists could expand the scope of resources (e.g., knowledge, data and expertise) they could access for producing novel ideas by two channels, i.e., first-time collaboration and international collaboration. The novel global challenge and the urgent need for effective vaccines might encourage the adjustment of team assembly towards effective teamwork that sparks new ideas by including newcomers beyond team members' pre-existing relationships and reaching out to international networks (Guimera, Uzzi, Spiro, & Amaral, 2005; Porac et al., 2004; Wagner, Whetsell, & Mukherjee, 2019). First-time collaboration indicates collaboration between two authors who have never collaborated with each other in the past so that scientists established collaboration outside their existing collaborative networks. First-time collaboration increases team freshness, facilitates scientists with wide reach (Gao, Yin, Myers, Lakhani, & Wang, 2021a) and helps acquire more complementary academic resources. Therefore, prior literature found that first-time collaboration it is positively related to research originality (Zeng, Fan, Di, Wang, & Havlin, 2021). International collaboration allows access to skills, knowledge and other resources used for research across national borders (Freeman, Ganguli, & Murciano-Goroff, 2014; Wagner et al., 2019). International collaboration influences scientific novelty in two opposite directions. On the one hand, the reach of an international network expands the "search space" of teams and thus leads to access to more novel ideas, which facilitates scientific novelty (Schilling & Green, 2011). Furthermore, variety and cross-cultural differences caused by international collaboration could contribute to greater creativity and high impacts (AlShebli, Rahwan, & Woon, 2018; Lee et al., 2015). However, international collaboration can also impede novelty due to higher transaction costs, communication barriers and audience effect (Wagner et al., 2019). Although researchers pointed out various barriers that impede international collaboration and first-time collaboration including an increasingly tense geopolitical climate (Lee & Haupt, 2021), the complexity of constructing new collaboration or collaborating internationally (Aviv-Reuven & Rosenfeld, 2021), physical and political obstacles (Cai et al., 2021), and high search and coordination costs (Fry et al., 2020b), we expect that these two types of collaboration might increase during the pandemic due to resource constraints and the urgent need for novel solutions to the disease. To investigate the possible mechanisms of changes in scientific novelty from the perspective of resource searching, we raise a question: **RQ2. How did first-time collaboration and international collaboration evolve during COVID-19?**

---

[1] https://www.nytimes.com/interactive/2020/science/coronavirus-vaccine-tracker.html
[2] https://oecd-opsi.org/covid-response/m
[3] 82.68% of coronavirus-related papers included in the COVID-19 Open Research Dataset are produced by non-single authors.

The aforementioned discussion suggests a potential association between first-time collaboration/international collaboration and scientific novelty, while whether their relationships were disrupted during COVID-19 remains unclear. The comprehensive influence of COVID-19 on the scientific community might reshape the benefits and detriments caused by the two types of collaboration that impact the generation of novel ideas, and distorted their association. The relationship between first-time/international collaboration and scientific novelty in the normal period might not hold for that during the COVID-19 period. Thus, we propose the third question: **RQ3. Is the relationship between first-time collaboration or international collaboration and scientific novelty during COVID-19 different from that in the normal period?**

We focus on the coronavirus-related domain as scientists in this field were most affected by COVID-19, which allows us to capture the immediate impact of COVID-19. One challenge in this study lies in measuring scientific novelty in the biomedical domain. Scientific novelty is conceptualized as a recombination of antecedent knowledge elements in an unusual fashion (Fleming, 2001; Kogut & Zander, 1992; Nelson & Winter, 1982; Weitzman, 1998). The combinatorial perspective of novelty was reflected in the process of generating COVID-19 solutions (Lee & Trimi, 2021). For example, the Draganfly's COVID-19 surveillance drone is a result of recombining several existing sensing technologies.[4] We follow the long-standing tradition of combinatorial novelty and measure novelty based on unusual combinations of preceding knowledge components. Bio-entities, such as genes, diseases and proteins, constitute the basic units of knowledge in the biomedical domain (Xu et al., 2020; Yu et al., 2021), and thus we use bio-entities to represent knowledge elements in coronavirus-related papers. We apply a cutting-edge word embedding technique, BioBERT (Bidirectional Encoder Representations from Transformers for Biomedical Text Mining) because it captures domain-specific information in the field of biomedicine. The advantages of our method are discussed in detail in Supplementary Note 1.

---

[4] https://cmr.berkeley.edu/2020/06/innovation-entrepreneurship/

Another challenge is how to measure scientific collaboration. Scientific collaboration is a process whereby researchers work together for a common goal of creating new scientific knowledge (Katz & Martin, 1997). A research team that is composed of a group of researchers could reflect scientific collaborative activities between group members (Stokols, Hall, Taylor, & Moser, 2008). Although co-authorship data fails to capture informal collaborative activities, such as sharing and exchanging data and ideas (Lewis, Ross, & Holden, 2012), researchers who are listed as authors on a research paper reflect a visible and easily quantifiable manifestation of collaborative efforts (Milojević, 2014). Co-authorship data allows the capturing of key elements of collaboration (Hara, Solomon, Kim, & Sonnenwald, 2003), and thus has been widely used as a reliable measurement of scientific collaboration (Leahey, 2016; Wuchty et al., 2007). Therefore, we consider a group of authors who appeared on a research article as a scientific team, based on which we identify the two types of collaboration patterns, first-time collaboration and international collaboration.

To address the research questions, we treat the outbreak of COVID-19 as a natural experiment and use a difference in differences (DID) approach to explore how scientific novelty, first-time collaboration and international collaboration evolved from January 2018 to December 2020. To address RQ1 and RQ2, in addition to exploring whether or not the occurrence of the COVID-19 is related to the changes in scientific novelty, and the two types of collaboration, we investigate whether the extent to which countries were affected by COVID-19 measured by the number of new cases and deaths, and outcome variables to strengthen the link between COVID-19 and outcome variables. We further explore the dynamic effect of COVID-19 on scientific novelty and the two collaboration patterns. To address RQ3, we investigate the relationship between first-time collaboration/international collaboration and scientific novelty before and during COVID-19.

This paper presents the first econometric study on the influence of COVID-19 on scientific novelty, which adds a new perspective that will be helpful to explain the evolution of scientific novelty in the presence of environmental threats. By investigating first-time collaboration and international collaboration, this study captures the shifted structure of scientific teams during COVID-19 that includes scientists' increasing preferences for constructing new collaboration links and their reluctance to collaborate internationally.

## Literature review
### The impact of crises on scientific collaboration

Existing literature shows the changes in the structure of scientific teams and narrower team membership during global crises, especially in the early stages of events. Early studies found that only 17% of SARS-related papers were internationally collaborated, which is lower than the average international collaboration rate in modern science (Chiu, Huang, & Ho, 2004). Recent studies have investigated the mixed pattern of international collaboration during COVID-19. Analyzing 10,432 coronavirus-related articles and preprints published from January 2020 to April 2020, Fry et al. (2020b) found smaller team size and fewer internationally collaborated papers during COVID-19, compared to the pre-pandemic times. Exploring COVID-19 related articles and non-COVID-19 related articles published in each of the first six months of 2016 to 2020, Aviv-Reuven and Rosenfeld (2021) found that international collaboration diversity that was measured by the number of collaborating countries was lower than in non-COVID-19 papers and than previous years. Similar findings of the shrink of international collaboration and team size were obtained in a few other studies (Homolak,

Kodvanj, & Virag, 2020; Wagner et al., 2021). The reduction in team size and international collaboration could be due to the declining time to spend on research (Myers et al., 2020), the complexity of conducting international studies (Aviv-Reuven & Rosenfeld, 2021), physical and political obstacles (Cai et al., 2021), and high costs for searching and coordination (Fry et al., 2020b).

In contrast, some studies provided empirical evidence that supports the promoting effect of crises on collaboration. Scientific collaboration was expected to increase because of the urgency to generate effective vaccines, the high-risk investment in anti-pandemic products by individual nations, and resources constraints for research during pandemics (Gates, 2020; Lee, Kang, & Kim, 2020). Focusing on the 2014 West African Ebola epidemic, Fry (2021) found increasing collaboration between the most affected countries and developed countries. The author attributed growing international collaboration to the need of sharing expertise, knowledge, data and other resources between local scientists and foreign scientists. Based on outbreaks of six infectious diseases, a study found that European countries and North America intensively collaborated with regions of the outbreaks (Zhang, Zhao, Sun, Huang, & Glänzel, 2020). Analyzing 8619 journal articles concerning eight emerging pathogens indexed in the Scopus database, a study found that international collaboration in research on these diseases was relatively high for most countries (Sweileh, 2017). Drawing on publication data in Scopus, Lee and Haupt (2021) found that the percentage of international collaboration on 3,401 COVID-19 articles published from January 2020 to early May 2020 reached 33.58%, which is higher than that for non-COVID-19 articles published in the same period, and higher than that for COVID-19 related articles published in the past five years. A descriptive statistical analysis of 18,875 articles on coronavirus indexed in Web of Science showed that the proportion of international collaboration is rising in most countries during the pandemic (Belli, Mugnaini, Baltà, & Abadal, 2020).

Distinctive studies showed variations in international collaboration during COVID-19. Prior literature found a consistent pattern that less-resourced, small and emerging countries in coronavirus research are more likely to be involved in international collaboration in the pandemic period than the major producers of coronavirus research (Belli et al., 2020). For example, China, the lead producer of COVID-19 studies, demonstrated a substantially lower rate of international collaboration (20%) on COVID-19 research, compared to that (22.48%) in the pre-pandemic times, and that for non-COVID-19 publications published in the same period (Lee & Haupt, 2021). Studies also showed that past international collaboration and the extent to which the country was affected by COVID-19 increases the odds of international collaboration on COVID-19 articles, while the country's relative wealth has a negative relationship with international collaboration on those articles during the pandemic (J. J. Lee & Haupt, 2021).

In summary, prior studies on collaboration during crises, indicated that scientists changed team formation to adapt to the needs of dealing with current issues. The inconsistent results concerning the question of whether crises impeded or enhanced collaboration stem from heterogeneity regarding disease and scientific domains investigated, databases and analysis techniques employed and the time frame of data. A majority of previous studies on scientific collaboration during COVID-19 or other crises focused on team size and international collaboration, mainly investigated the early stage of events, performed bibliometric analyses with a few exceptions. It is still unclear how other types of collaboration, such as first-time collaboration, evolved during COVID-19, and how changes in collaboration influenced team performance, such as scientific novelty.

**The impact of crises on innovation and scientific novelty**

Prior studies have focused on crises that have been external drivers of innovation, such as the 9/11 terrorist attacks (Birkland, 2004), financial crises (Knudsen, 2019) and COVID-19. To support the sluggish economy, and stimulate research and development for the pandemic, innovative products, ideas and services are required during COVID-19 (Ramadi & Nguyen, 2021). Conventional approaches are not sufficient to fully address pandemic-related problems. Many economists criticized existing growth and innovation models, and viewed COVID-19 as a good opportunity for the exploration of alternative innovation and growth models (Wu & Sheikh, 2021). In such an extraordinary circumstance, governments, research institutions, industries and even individuals were turned into problem solvers and tried to generate innovative ideas to tackle a common adversary (Patrucco, Trabucchi, Frattini, & Lynch, 2021; Ramadi & Nguyen, 2021). Governments launched large-scale, fast-tracked and top-down innovation initiatives and policies to spur new technologies and solutions, including hackathons and financial support (Patrucco et al., 2021; Ramadi & Nguyen, 2021). Open innovation, sustainable innovation, crowdsourcing innovation and frugal innovation were considered highly effective in addressing multi-faced problems caused by COVID-19 (Dubey, Bryde, Foropon, Tiwari, & Gunasekaran, 2021; Patrucco et al., 2021; Ramadi & Nguyen, 2021; Sarkis, 2020). Based on survey data on 237 knowledge workers in Norway, a study found that the increased use of digital platforms improved individuals' creative performance in the context of work from home caused by COVID-19 (Tønnessen, Dhir, & Flåten, 2021).

Researchers demonstrated that innovation, especially pandemic-related one, has been improved during COVID-19 through the invention of new ideas, improvement of applications and implementation of new technologies. Forced by the global public health emergency, innovation processes that usually took years in the pre-pandemic period have turned more radical, and even big achievements in science and technology could be obtained in days (Brem, Viardot, & Nylund, 2021). For instance, it only took 69 days for the first COVID-19 vaccine to reach the human trial stage after the identification of the causative agent of COVID-19, which is far shorter than 25 months spent for the outbreak of SARS (Kim et al., 2020). Usefulness and applications of emerging technologies have been substantially accelerated during COVID-19 due to their important roles in affected sectors of the pandemic, such as 3D printing, big data analytics, distance education and blockchain (Brem et al., 2021; Farah, 2018). Investigating 3,001 A-share listed companies in China, a study found that the severity of COVID-19 measured by the number of days of the first-level public health emergency response initiated by each province, is significantly and positively related to enterprises' innovation performance proxied by the R&D investment (Han & Qian, 2020).

Previous studies suggested the keys to promoting innovation in COVID-19, including convergence innovation, repurposing existing knowledge, and access to essential resources through collaboration. Lee and Trimi (2021) developed the concept of convergence innovation that indicates the combination of the various technologies, ideas and strategies, and argued that convergence innovation could be a catalyst for managing COVID-19. The pandemic stimulated the innovativeness of many companies that repurposed their slack and created product innovation (Gopalakrishnan & Kovoor-Misra, 2021). Manufacturers, such as General Motors and Dyson, used their manufacturing capacities to create ventilators that were in short supply.[5] Analyzing 350 applications for two competitions, a study found that COVID-19 has revolutionized the way where innovative solutions are generated (El Akoum & El Achi, 2021b). That is, repurposing technologies and ideas could be effective and cost-efficient for generating solutions to complex problems. Based on 185 small and medium-sized companies in Iran, a study found that collaboration led to greater innovation during COVID-19 (Van Auken,

---

[5] https://www.mckinsey.com/business-functions/strategy-and-corporate-finance/our-insights/innovation-in-a-crisis-why-it-is-more-critical-than-ever

Ardakani, Carraher, & Avorgani, 2021). During COVID-19, because of a lack of adequate resources, some countries were not able to meet the traditional processes of testing and trailing new drugs and technologies (Dubey et al., 2021), which forced them to collaborate with other countries for sharing information and other necessary resources, such as the DNA of the original virus and infection patterns (Lee & Trimi, 2021). Collaboration was also formed among scientists, firms, governments and universities to develop effective and innovative vaccines for COVID-19 (Desmond-Hellmann, 2020).

The current studies provided multiple perspectives for understanding innovation processes during COVID-19 and emphasize the importance of collaboration for the development of innovation, while empirical efforts that supported the improvement of innovation were limited, as well as how the evolution of innovation and scientific novelty is related to changes in collaboration. To address the research gaps, this study explores the evolution of scientific novelty in the pandemic, as well as the mechanisms of such changes by focusing on two types of collaboration that are important for searching complementary resources, first-time collaboration and international collaboration.

## Data

Two major datasets are used in this study, with one including publication data on coronavirus research that is used to measure an individual paper's scientific novelty and capture authors' country information, and the other including country-by-country patient data about COVID-19 that is used to identify the timing when the first COVID-19 case was confirmed in a country. Publication data on coronavirus research is collected from the COVID-19 Open Research Dataset [6] (hereafter CORD-19) that covers research articles about COVID-19 and related historical coronaviruses, e.g., SARS and MERS, that were published before December 2020. This dataset was downloaded on August 9$^{th}$, 2021. It includes title, abstract, author name, DOI, PubMed ID, and publication date. CORD-19 papers are sourced from PubMed Central, bioRxiv and medRxiv, with title, abstract or full text including the following keywords: "COVID-19" OR "Coronavirus" OR "Corona virus" OR "2019-nCoV" OR "SARS CoV" OR "MERS-CoV" OR "Severe Acute Respiratory Syndrome" OR "Middle East Respiratory Syndrome". The distribution of papers per year in CORD-19 (Fig.S1a) indicates a sudden growth of papers in the years of significant pandemics.

We use the patient data on COVID-19 derived from the website of Our World in Data that covers 211 countries from December 2019 to December 2020,[7] to capture the timing when the first COVID-19 case was officially confirmed, and the daily number of new COVID-19 cases and deaths in each sampled country during the December 2019-December 2020 period. The distribution of COVID-19 cases and deaths in each month is illustrated in Fig.S2.

We identify authors' country information based on the 29 million PubMed dataset that covers 1800-2020 with author names disambiguated.[8] Based on DOI and PubMed ID provided in the CORD-19 dataset, 204,936 CORD-19 papers are linked to their versions in the PubMed dataset and thus the following information of CORD-19 papers was obtained: authors' unique identifiers and authors' address information. Authors' unique identifiers allow us to know whether authors in a paper have collaborated in the past according to their publications records in the PubMed database, which enables us to identify first-time collaboration. Authors' affiliation information helps us to identify country names from authors' address information in each article by manually merging variations (e.g., ISO two-letter or three-letter country codes, alternative country names, and country names with typos) of country names into the same country. Finally, standard country names corresponding to authors' locations in 164,485 CORD- papers that include 288,303 unique authors are found.

---

[6] Accessible at https://www.kaggle.com/allen-institute-for-ai/CORD-19-research-challenge,
[7] Accessible at https://ourworldindata.org/covid-cases.This dataset was downloaded on September,2021.
[8] Author name disambiguation (AND) in PubMed has achieved through the integration of two existing AND datasets: Authority and Semantic Scholar. The precision of AND of PubMed was evaluated using the NIH ExPORTE-provided information on NIH-funded researchers. The evaluation results show that AND in PubMed achieved an F1 score of 98.09%. The algorithms of the PubMed dataset's author name disambiguation and the validation are shown in a recent study (Xu, Kim, Song et al., 2020).

The final dataset used for the regression analysis includes 98,981 research articles published from January 2018 to December 2020 by the top 50 prolific countries that are ranked by the number of coronavirus-related papers published during the study period. To measure the country's productivity, we use a full counting method (Waltman, 2016) based on the authors' country information. [9] The top 50 prolific countries regarding the productivity of coronavirus research during the study period are selected as the sampled countries. The productivity of the 50 sampled countries/regions is shown in Table S1. The distribution of CORD-19 papers by month and country from January 2018 to December 2020 is indicated in Fig.S3.

# Methods

### Measuring scientific novelty of papers

133,590 unique bio-entities were extracted from 164,100 CORD-19 papers' titles and abstracts using PubTator central (detailed in Supplementary Note 2). To measure the novelty of entity combination of CORD-19 papers, we use BioBERT to capture the distance between two bio-entities in each entity pair. The introduction to BioBERT and the reasons why we use it to generate embeddings of bio-entities are shown in Supplementary Note 3.

The pipeline of generating sub-word representation for each bio-entity extracted from CORD-19 papers using BioBERT is indicated in Fig.S5. Based on the embedding of bio-entities extracted from CORD-19 papers, we calculate the cosine distance defined in Equation 1 between two resulting vectors corresponding to each entity in an entity pair.

$$cosine\ distance_{i,j} = 1 - \frac{i \cdot j}{\|i\|_2 \|j\|_2} \quad (1)$$

where $i$ and $j$ indicate two entities in an entity pair; $i \cdot j$ refers to the dot product of $i$ and $j$; $\|i\|_2 \|j\|_2$ means the product of $i$'s and $j$'s Euclidean norm.

We extract 133,590 unique bio-entities using Pubtator Central from titles and abstracts of CORD-19 papers published in and before December 2020 and pair them up. The cosine distance of two entities in each of 783,442 entity pairs detected in CORD-19 paper is captured from the resulting embedding using BioBERT. We examine the distribution of the distance between two entities in entity pairs extracted from CORD-19 papers (Fig.S6a), and consider an entity pair in which the distance of two entities is in the upper 10[th] percentile of this distribution as a novel entity combination. The novelty score for each paper is measured by the proportion of novel entity pairs according to our definition of novelty entity combination to the possible number of entity pairs in a paper. The formula used to calculate the novelty score for a paper is shown as:

$$Novelty\ score_i = \frac{m}{C_n^2} \quad (2)$$

where $i$ denotes paper $i$; $n$ indicates the number of bio-entities extracted from paper $i$;

---

[9] For example, for a paper authored by two scientists with Chinese affiliations, one scientist with a US affiliation and three scientists with UK affiliations, China, the US and the UK get one paper, respectively. Hence, overall three publications are allocated to these three countries. For scientists with several affiliations belonging to different countries, we take into account the country information of the affiliation listed as the first.

$C_n^2$ refers to the number of combinations of two that can be drawn from the set of $n$ bio-entities extracted from paper $i$, i.e., the number of entity pairs generated by $n$ bio-entities; $m$ denotes the number of entity pairs in which two entities' distance is in the upper $10^{th}$ percentile of the distribution of the distance of two entities in all entity pairs generated from CORD-19 papers. For example, for a paper that contains three bio-entities (i.e., entity $a$, $b$ and $c$), the number of entity pairs for this paper is three. If the distance between $a$ and $b$ is in the upper $10^{th}$ percentile of the distribution (Fig.S6a), the novelty score for this paper is 1/3. The higher the novelty score, the more novel entity combination in a paper.

**Variables**

To address RQ1 and RQ2, the major independent variable is whether the first case of COVID-19 (*COVID19*) has been confirmed in the country by the month. We identify the month when the first case of COVID-19 was officially confirmed in each of the 50 sampled countries based on the patient data. Once the first COVID-19 case has been confirmed in the country, the country gets treated in the month and the succeeding months. The distribution of treated countries, i.e., the countries where the first COVID-19 case has been confirmed, and untreated countries, i.e., those where the first COVID-19 case has not been confirmed, by the month is indicated in Fig.S7.

<u>Paper-level variables</u>

To address RQ2, we measure first-time collaboration ratio and international collaboration. International collaboration for a paper is a binary variable that is determined by whether authors listed in a paper are from at least two countries. It is one if at least two authors are from different countries, and zero otherwise. First-time collaboration ratio for a paper indicates the fraction of author pairs where two authors did not collaborate in the past based on authors' publication history in the PubMed database[10] to the total number of author pairs in a paper, measuring the degree to which first-time collaboration is involved in the team, which is defined in Equation 3

$$First-time\ collaboration\ ratio_i = \frac{p}{C_n^2} \quad (3)$$

where $i$ denotes paper $i$; $C_n^2$ refers to the number of combinations of two that can be drawn from the set of $n$ authors listed in paper $i$; $p$ indicates the number of author pairs in which two authors have no prior collaboration. The higher the first-time collaboration ratio for a paper, the more first-time collaboration involved in the team of the paper.

Team size is included as a control variable to address the three RQs since it could influence scientific novelty (Hülsheger, Anderson, & Salgado, 2009; Lee et al., 2015) and it might be related to international collaboration and first-time collaboration (Gao et al., 2021b). We use the number of co-authors in a paper to measure team size at the paper level.

<u>Country-level variables</u>

All paper-level variables were aggregated to the country level for country-level analyses. Using a full counting method, we use an example to demonstrate how paper-level variables are calculated to country-level variables as shown in Fig.S8.

---

[10] Authors might have prior collaboration in the domains that are not included in the PubMed database, which is not captured in this study.

In RQ1 and RQ3, the dependent variable is a country's average novelty score (*novelty score*) of entity combination for papers published by this country in a given month, which quantifies the monthly average extent to which entities are combined rarely for knowledge production of the country. The higher the novelty score, the more novel countries' knowledge production in a month. Team size is considered an influential factor of scientific novelty. We use the country's monthly average number of authors in CORD-19 papers to measure the average team size (*team size*) of coronavirus papers in a country, as a control variable to estimate the country's average novelty score for addressing RQ1 and RQ3.

In RQ2, the dependent variables include: (1) the proportion of internationally collaborative papers in a country in a given month, which is used to reflect the degree to which the papers are internationally collaborative (*international collaboration ratio*); (2) the country's average first-time collaboration ratio (*first-time collaboration ratio*), which is used to measure the extent to which first-time collaboration is involved in teams for CORD-19 papers published in the month. In RQ3, we explore the association between first-time collaboration ratio/ international collaboration ratio and scientific novelty so that these two variables are explanatory variables in RQ3.

In RQ1 and RQ2, in addition to exploring whether or not the occurrence of COVID-19 influenced scientific novelty, and the two types of collaboration, we examine whether the severity of COVID-19 is related to changes in countries' scientific novelty and those in the two collaboration patterns, to further confirm the link between COVID-19 and dependent variables. We use the number of new COVID-19 cases and deaths per million to measure the degree to which countries were affected by COVID-19. The daily numbers of new COVID-19 cases (*COVID19 case*) and deaths (*COVID19 death*) per million confirmed in each sampled country are aggregated to the month level and considered as two explanatory variables.[11]

Summary statistics of variables and the correlation matrix across variables are shown in Tables S2 and S3, respectively.

**Difference in differences (DID)**

To address RQ1 and RQ2, we use the outbreak of COVID-19 as a natural experiment to explore how scientific novelty, and first-time collaboration and international collaboration, evolve before and during the pandemic by using a DID approach. The analyses are performed based on the data on 50 sampled countries over 36 months from 2018 January to 2020 December. In RQ1, we regress the dependent variable, i.e., *novelty score,* on whether the first case of COVID-19 in the country (*COVID19*) has been confirmed by the month and other covariates that might influence scientific novelty as shown in Equation 4. We apply an OLS linear model that contains fixed effects for country, $\theta_i$, and those for month, $\delta_t$, to control the time-invariant and country-invariant factors. The coefficient on *COVID19* is a before-after estimate of the impact of the pandemic on scientific novelty.

$$Novelty\ score_{i,t} = \alpha + \beta COVID19_{i,t} + \gamma control_{i,t} + \delta_t + \theta_i + \varepsilon \quad (4)$$

Similarly, to address RQ2, we use the DID strategy to investigate the association between the countries' first-time collaboration ratio in the month/the fraction of internationally collaborative papers by the country in the month and the outbreak of the COVID-19 in the country. The fixed effects of countries and months are included.

---

[11] The numbers of COVID-19 deaths and cases are not normalized by the country's population as country-fixed effect has been included to estimates scientific novelty. Country-fixed effect is used to control any country-invariant factor including country's population that cannot change a lot during a short period.

To solve RQ1 and RQ2, we further explore the relationship between the severity of COVID-19 in the country and the country's novelty score, we regress the country's average novelty score in the month on the monthly number of new COVID-19 cases and deaths. Control variables are the same as those in Equation 4. The details of investigating the dynamic effects of COVID-19 on scientific novelty and the two collaboration patterns are shown in Supplementary Note 4.

**Regression**

We address RQ3 by conducting regression analyses including interaction terms and sub-sample analyses. Papers' novelty scores are estimated by Equation 5:

$$Novelty\ score_i = \alpha + \beta_1\ COVID19 + \beta_2\ first-time\ collaboration + \beta_3\ international\ collaboration + \beta_4\ first\_C \times COVID19 + \beta_5\ international\_C \times COVID19 + \beta_6 team\ size + \delta_t + \varepsilon\ (5)$$

where *i* denotes a paper; novelty score indicates the proportion of entity pairs that are highly distant to the possible entity pairs in a paper; *COVID19* is a binary variable that is one if the paper is published in and after December 2019, and zero otherwise; *first_collaboration* indicates the proportion of author pairs in which two authors have no prior collaboration in the past to the possible author pairs in a paper; *international_collaboration* is a binary variable that is one if the team includes authors from at least two countries, and zero otherwise; *team size* is a control variable that indicates the number of authors listed in a paper; fixed effects regarding the papers' publication year ($\delta_t$) are included; to explore the relationship between the two collaboration patterns and the papers' novelty score before and during COVID19, the interaction terms between the two collaboration patterns and the occurrence of the outbreak of COVID-19 are introduced to the model, i.e., $first\_C \times COVID19$, and $international\_C \times COVID19$.

Sub-sample analyses are conducted to confirm the relationship between the papers' novelty score and the two collaboration patterns before and during the pandemic by separating all coronavirus papers into two groups, with papers published before December 2019, and those published after that month. Then, we estimate the relationship between papers' novelty and the two collaboration patterns based on these two groups of papers, separately.

We use an existing approach proposed by Azoulay et al. (2011) to measure scientific novelty and perform our analyses again for robustness check (Supplementary Note 5). Generally, the major findings are consistent.

# Results

For RQ1, our findings suggest that coronavirus research has become more novel since the outbreak of COVID-19. After 2019, the year of the COVID-19 outbreak, there was a dramatic increase in the average novelty score of global coronavirus research relative to the earlier years (see Fig.1a). Since the global first COVID-19 case was officially confirmed in December 2019, the average novelty score of global coronavirus papers sharply increased until April 2020, then slightly declined and remained stable (see Fig.1b). The results of the DID regression show that "treated" countries (i.e., countries with an infection) have a 0.048 (p-value<0.01) higher novelty score than "untreated" countries (i.e., countries without infection)—this is an increase of 53.15 % standard deviation (see column 1 in Table S4). The estimated dynamic impact of a COVID-19 outbreak in a country on the country's scientific novelty score of coronavirus literature is shown in Fig.2a and column 2 of Table S4, which illustrates a jump in countries' average novelty scores in the first month (i.e., t+1 where t refers to the month the first COVID-19 case was confirmed in a country) after the occurrence of the first COVID-19 case in a country, while there is no significant difference between treated and untreated countries before the first COVID-19 case in the country. Furthermore, the regression results show that more COVID-19 cases (coefficient: 0.006, p-value<0.01) and deaths (coefficient: 0.010, p-value<0.01) per million in a month predict a higher scientific novelty (columns 1 and 2 of Table S5), suggesting that the increased scientific novelty is associated with the severity of the local outbreak.

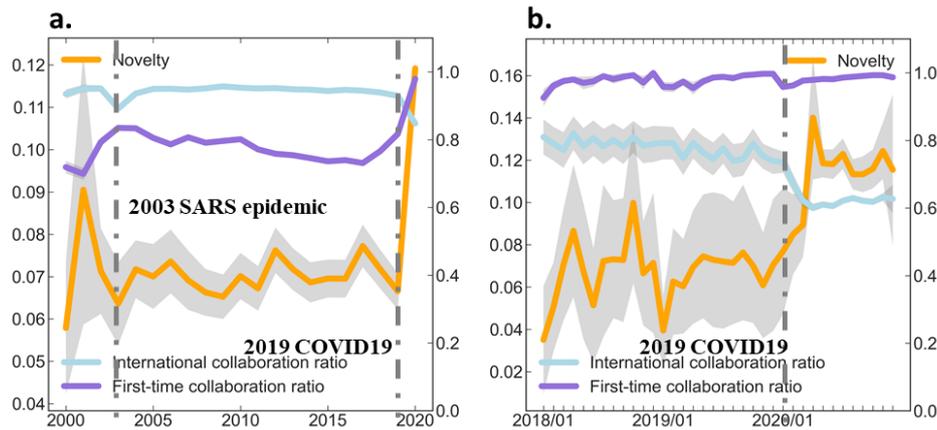

**Fig 1. The trend of average novelty score and first-time/international collaboration ratio for global coronavirus research.** The left vertical axis in each sub-figure indicates the novelty score of papers and the right one refers to first-time/international collaboration ratio. International collaboration ratio indicates the proportion of internationally collaborative papers, and first-time collaboration ratio refers to the fraction of first-time collaboration defined as collaboration between two authors without prior collaboration in scientific teams. In sub-figure b, the study period is from January 2018 to December 2020, a total of 36 months. The shaded areas represent upper and lower bounds of 95% CIs.

After the global first COVID-19 case, we found that international collaboration declined and first-time collaboration increased. Fig.1b presents the sudden decrease in global coronavirus papers' international collaboration ratio after December 2019. After its lowest point, the international collaboration ratio showed a steady trend at the level of 0.6. DID estimates suggest that a country's proportion of internationally collaborative papers in coronavirus research shrank by 6.3% (coefficient: -0.063, p-value < 0.01 in column 3 in Table S4) after the occurrence of the first COVID-19 case in the country. We further find that first-time collaboration ratio increased by 0.7% (coefficient: 0.007, p-value < 0.01 in column 5 in Table S4) after the report of the first COVID-19 case in a country, which is an increase of 18.9% standard deviation. This suggests that after the first case is confirmed in a country, more first-time collaboration is found in coronavirus research for that country. Furthermore, the monthly number of new COVID-19 cases and that of new deaths per million are both significantly negatively related to the international collaboration ratio in the country (see columns 3 and 4 in Table S5). The dynamic impact of the first COVID-19 case in a country on its average first-time collaboration and international collaboration ratio is estimated in columns 4 and 6 in Table S4, respectively, and is illustrated in Fig.2b and c. It should be noted that first-time collaboration ratio increased only in the month of the outbreak (coefficient: 0.013, p-value<0.01), and in the first (coefficient: 0.011, p-value<0.05) and second (coefficient: 0.010, p-value<0.05) months since the outbreak. The impact of COVID-19 on first-time collaboration only occurred during the early stage of the pandemic and was not long-lasting.

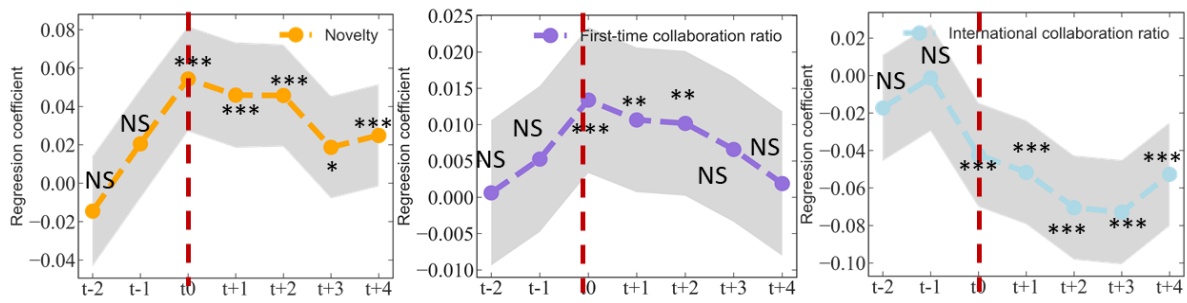

**Fig 2. The DID estimates of the relationship between the occurrence of the first case of COVID-19 in the country and the countries' average novelty scores, first-time collaboration ratio and international collaboration ratio in a month.** International collaboration ratio indicates the proportion of internationally collaborative papers by the country in a month, and first-time collaboration ratio refers to the fraction of first-time collaboration defined as collaboration between two authors without prior collaboration in scientific teams by the country in a month. T-n indicates n month(s) before the month (t0) when the first COVID-19 case was confirmed in the country, and t+n indicates n month(s) after t0. ***, ** and * represent significance at the 1%, 5%, and 10% level. The shaded areas represent upper and lower bounds of 95% CIs.

There is a sudden change in scientific novelty, international and first-time collaboration ratio around the year of the outbreak of SARS, with the same direction we find during the COVID-19 outbreak (see Fig.1a). This might suggest that the pattern we observed can be generalizable during the pandemic period.

We further find that papers with a higher first-time collaboration ratio are predicted to be more novel during the pandemic. Moreover, international collaboration did not hamper scientific novelty during the pandemic as it had done in the pre-pandemic period. Fig.3 and columns 1 and 2 in Table S6 illustrate papers' predicted novelty score estimated by a regression model including interaction terms between papers' first-time collaboration ratio or international collaboration and the occurrence of the first global COVID-19 case. It suggests that before COVID-19, papers' first-time collaboration ratio was significantly negatively related to papers' novelty scores. However, this relationship turned significantly positive for papers published during the COVID-19. Before COVID-19, international collaboration of papers was significantly negatively associated with papers' scientific novelty. The relationship became positive although it was not significant during COVID-19, which means that international collaboration has not impeded scientific novelty since the occurrence of COVID-19 (see Fig.3). The subsample analyses also confirm these findings (see columns 3 and 4 in Table S6).

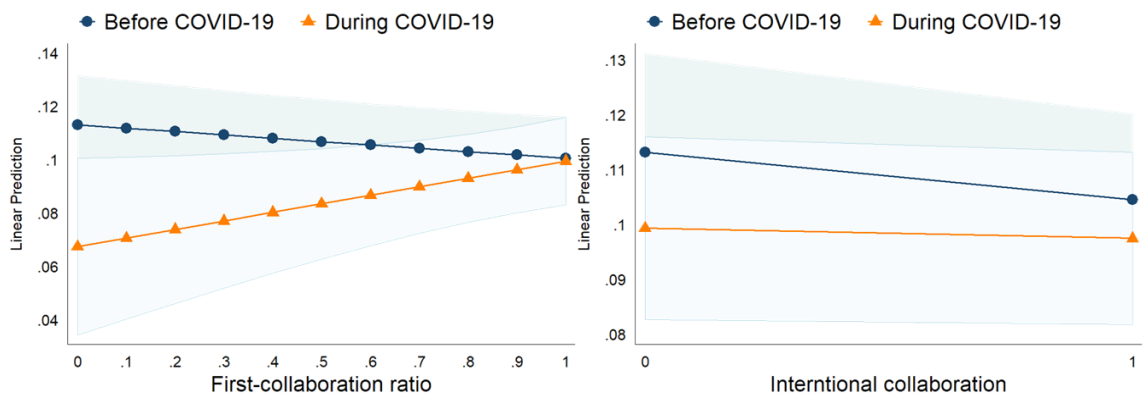

**Fig. 3. The linear prediction of papers' novelty score before and during COVID-19.** The X-axis in sub-figure a and b indicates the level of papers' first-time collaboration ratio and whether the paper is internationally collaborative, respectively. The Y-axis indicates the predicted novelty scores of papers for levels of variables in the X-axis before (the orange line) and during COVID-19 (the blue line) when all other covariates are set to their means. The shaded areas represent upper and lower bounds of 95% CIs.

## Discussion and Conclusion

Our results show that in the initial period following a coronavirus outbreak, scientific novelty dramatically increased, which suggests scientists' efforts to try novel recombinations of existing knowledge to combat this global pandemic. The fraction of first-time collaboration, i.e., collaboration between team members without prior collaboration, in scientific teams engaged in coronavirus research grew, and the proportion of internationally collaborative papers sharply decreased.

In the pre-COVID19 period, first-time collaboration is significantly negatively associated with a paper's novelty score, while this relationship turns significantly positively related to a paper's novelty during the pandemic. From the psychological perspective, as opposed to repeat collaboration, first-time collaboration leads to high costs of adaption and socialization (Chen, 2005; Kammeyer-Mueller, Wanberg, Rubenstein, & Song, 2013), communication and coordination (Petersen, 2015), and less trust and more unfamiliarity (Rockett & Okhuysen, 2002; Van Der Vegt, Bunderson, & Kuipers, 2010), which might dampen scientific novelty in normal times (Fry et al., 2020a; Granovetter, 1985; Guimera et al., 2005). During COVID-19, efficiency and speed were substantially emphasized due to the urgent need to generate novel solutions. When constructing new collaboration with others for acquiring complementary resources, scientists might try their best to overcome ineffective communication and coordination, which might mitigate the detrimental effects of first-time collaboration. First-time collaboration allows the pooling together of a broader scope of information, data and resources outside the preexisting relationships and conflicts that might improve scientific novelty (Porac et al., 2004; Skilton & Dooley, 2010; Yong, Sauer, & Mannix, 2014). The benefits of first-time collaboration might outweigh its disadvantages that have been reduced during the pandemic. Therefore, we observe that during the pandemic, papers produced by teams with a larger proportion of first-time collaboration are more novel.

We find that there is insignificant difference in novelty scores between internationally collaborative papers and their counterparts during the pandemic. However, we observe a negative association between international collaboration and papers' novelty in the pre-pandemic period. This negative relationship was also supported in a recent paper by Wagener et al. (2019) who suggested that international collaboration produces less novel and more conventional knowledge combinations based on data extracted from Web of Science and Scopus in 2005. Wagener and her colleagues attributed this negative relationship to various factors, such as higher transaction costs of international research that impede high novelty (Lauto & Valentin, 2013; Ou, Varriale, & Tsui, 2012), dependence on information technologies and English that might limit effective communication (Lagerström & Andersson, 2003), and audience effect. Global pandemics normally emerged in a few regions and spread on a global scale, which makes data on pandemics locally distributed. International collaboration allows the timely exchange of data, genetic and viral material, and other complementary resources across national borders, which is essential for accelerating the development of cures. Furthermore, to tackle health emergencies, scientists in the field spared no effort to conduct COVID-19-related research. The increasing efforts of scientists could offset the costs of communication and coordination caused by international collaboration that were considered the major detriments of novelty (Wagner et al., 2019). In this case, international collaboration might not hamper as it does during normal periods. The examples could be the generation of successful COVID-19 vaccines through international collaboration.[12]

With rapidly developing globalization and the increasing complexity of economic, societal, political, and environmental issues, the traditional perception of normal science with the assumption that the research system operates with institutional stability (Kuhn, 1962), is no longer sufficient to address issues or problems in the scientific community. Local and even global research systems could be immediately influenced by exogenous and unexpected events (Mryglod, Holovatch, Kenna, & Berche, 2016). This study provides evidence on how science progresses differently during a pandemic from a normal science period.

There are several limitations in this study. We only analyze coronavirus-related papers and do not include research papers in other fields and thus the findings of this study might not apply to other domains. Further investigation should be conducted for research papers in other domains, especially for those in humanities and social sciences. This study only includes publication data up to December 2020 because name disambiguation for research articles published in 2021 indexed by PubMed has not yet been processed. More recent publications should be analyzed for future study. In addition, first-time collaboration depends on whether two authors collaborated with each other in the past based on their publication history in the PubMed database. If their co-authored papers are not indexed by PubMed, their first-time collaboration will be overestimated in this study. The diversity of research institutions could shed light on collaboration at the institutional level, we will explore affiliations to see whether the team collaboration from different institutions will show different patterns in first-time collaboration and international collaboration for future study.

**Supplementary Information**

Supplementary Information (SI) is available for this paper:

https://portland-my.sharepoint.com/:w:/g/personal/weipingli4-c_my_cityu_edu_hk/EYUowjNh-B5DsEgMmGcoFJwBAF_grTTP-FwDrqPS8xfvRw?e=WZZFdg

---

[12] https://www.oecd-ilibrary.org/sites/e0643f52-en/index.html?itemId=/content/component/e0643f52-en#chapter-d1e10342